\documentclass[pdflatex,sn-mathphys]{template_files/sn-jnl}% Math and Physical Sciences Reference Style
\jyear{2021}%

\theoremstyle{thmstyleone}%
\newtheorem{theorem}{Theorem}%  meant for continuous numbers
\newtheorem{proposition}[theorem]{Proposition}% 

\theoremstyle{thmstyletwo}%
\newtheorem{example}{Example}%
\newtheorem{remark}{Remark}%

\theoremstyle{thmstylethree}%
\newtheorem{definition}{Definition}%

\newcommand{\boldface}[1]{\boldsymbol{#1}}  % italic (slanted)

\newcommand{\bfc}{\boldface{c}}

\newcommand{\bfx}{\boldface{x}}

\newcommand{\bfI}{\boldface{I}}

\newcommand{\calK}{\mathcal{K}}
\newcommand{\calL}{\mathcal{L}}

\newcommand{\calX}{\mathcal{X}}

\newlength{\boxwidth}
\def\btheorem{\begin{theorem}}
\def\etheorem{\end{theorem}}
\def\blemma{\begin{lemma}}
\def\elemma{\end{lemma}}
\def\bproposition{\begin{proposition}}
\def\eproposition{\end{proposition}}
\def\bcorollary{\begin{corollary}}
\def\ecorollary{\end{corollary}}
\def\bdefinition{\begin{definition}}
\def\edefinition{\end{definition}}
\def\bexample{\begin{example}}
\def\eexample{\end{example}}
\def\bremark{\begin{remark}}
\def\eremark{\end{remark}}

\newcommand{\be}{\begin{equation}}
\newcommand{\ee}{\end{equation}}

\newcommand{\beq}{\begin{eqnarray*}}
\newcommand{\eeq}{\end{eqnarray*}}
\newcommand{\bem}{\begin{multline}}
\newcommand{\eem}{\end{multline}}
\newcommand{\ba}{\begin{align*}}
\newcommand{\ea}{\end{align*}}

\renewcommand{\figurename}{Figure}
\renewcommand{\tablename}{Table}

\begin{document}

\title[Preprint]{Inverse-design of nonlinear mechanical~metamaterials via video~denoising~diffusion~models}

\author[1]{\fnm{Jan-Hendrik} \sur{Bastek}}\email{jbastek@ethz.ch}
\author*[1]{\fnm{Dennis M.} \sur{Kochmann}}\email{dmk@ethz.ch}

\affil*[1]{\orgdiv{Mechanics \& Materials Laboratory}, \orgname{Department of Mechanical and Process Engineering, ETH Zürich}, \orgaddress{
\city{8092 Zürich},
%\postcode{8092},
%\state{Zürich},
\country{Switzerland}}}

\abstract{The accelerated inverse design of complex material properties---such as identifying a material with a given stress-strain response over a nonlinear deformation path---holds great potential for addressing challenges from soft robotics to biomedical implants and impact mitigation. While machine learning models have provided such inverse mappings, they are typically restricted to linear target properties such as stiffness. To tailor the nonlinear response, we here show that video diffusion generative models trained on full-field data of periodic stochastic cellular structures can successfully predict and tune their nonlinear deformation and stress response under compression in the large-strain regime, including buckling and contact. Unlike commonly encountered black-box models, our framework intrinsically provides an estimate of the expected deformation path, including the full-field internal stress distribution closely agreeing with finite element simulations. This work has thus the potential to simplify and accelerate the identification of materials with complex target performance.}


\maketitle

\section{Introduction}
\label{sec:introduction}

Creating materials with tailored properties has gained popularity across disciplines ever since additive manufacturing enabled the manipulation of multi-material and cellular architectures across scales. Instead of choosing from the limited catalog of natural materials, engineers and designers now have access to the drastically expanded design and property spaces of so-called \textit{metamaterials}, which have been designed, among others, to achieve mechanical properties previously not attainable. Realizations of metamaterials have various forms, most commonly involving the periodic arrangements of small-scale structural building blocks.\cite{Tancogne-Dejean2018, Kumar2020, Kadic2019}

The physical mechanisms governing the mechanical behavior of such architected materials are mostly well understood, and various numerical frameworks such as the finite element (FE) method provide accurate structure-to-property relations, predicting the effective material properties based on an underlying small-scale architecture. By contrast, the inverse problem of identifying possible small-scale designs yielding a desired property has remained a challenge. Methods to address the latter include topology optimization (TO)\cite{Wu2021, Telgen2022, DaSilva2022} and, more recently, data-driven algorithms. Most of these approaches have, however, been restricted to \textit{linear} material properties such as the effective elastic stiffness in three dimensions (3D)\cite{Bastek2022,Zheng2021} or Poisson's ratio.\cite{Tian2022} Extensions to nonlinearity (e.g., via multi-material configurations) have been presented recently\cite{Li2022} but involve computationally expensive simulations.

While tuning a material's stiffness is sufficient for applications involving small deformation (such as patient-specific bone implants matching the native bone properties, or vibration insulation by attenuating linear waves), controlling the nonlinear response of soft metamaterials over a finite deformation path can unlock advanced functionality for emerging fields such as soft robotics\cite{Elango2015}, tissue engineering\cite{Chan2008}, and impact energy absorption.\cite{Zhang2020} Metamaterials with tailored stress-strain responses can, e.g., mimic the nonlinear response of human fingers\cite{Hyun-YongHan}, enable actuation of soft robots via ``snap-through instabilities''\cite{Yang2015}, or serve as biomimetic scaffolds assisting in artery restoration.\cite{Niu2019}

Unfortunately, the nonlinear setting drastically adds to the complexity of the (inverse) map from property to structure. Extensions of TO to nonlinear properties exist\cite{Wang2014,Medina2023} but remain challenging due to strong dependence on the initial guess and discretization\cite{Buhl2000}, lack of physical effects such as contact\cite{Xue2022}, and degrading solver stability when considering non-trivial mechanisms such as post-buckling.\cite{Huang2022} Most importantly, a single optimization study may require hours of runtime, which is a prime reason why recent studies focused on rather simple design spaces and optimization objectives.\cite{Abdi2018,Kim2020}

Over the past decade, the rise of deep-learning models with their unparalleled ability to identify highly nonlinear maps has presented a potential alternative. When applied to nonlinear material property prediction, deep learning has served as an efficient forward approximation (replacing costly FE simulations) in combination with genetic algorithms to iteratively identify structures with a target nonlinear response, including, e.g., shell-like metamaterials and quadrilateral structures.\cite{Wang2022, Deng2022} However, the considered design spaces have remained limited and predictions may lack physical intuition and rely on costly FE simulations to validate up to a hundred generated designs and to select the one closest to the desired stress-strain response.\cite{Deng2022}

These challenges resemble those addressed recently in the image generation community by (video) diffusion models. Diffusion models\cite{Sohl-Dickstein2015} have gained attention due to their ability to generate seemingly photo-realistic images based on text descriptors, a famous representative being DALL-E 2.\cite{Ramesh2022}, and have recently been extended to generate short video sequences with remarkable results.\cite{Ho2022} Compared to variational autoencoders\cite{Kingma2014} or generative adversarial networks\cite{Goodfellow2014}, diffusion models offer improved sample quality and simpler and more stable training protocols by gradually removing noise from a sample drawn from a prior distribution (typically unit Gaussian).\cite{Dhariwal2021}

The shift from linear to nonlinear material properties can, at a high level, be compared to going from image to video generation. In both cases, a new data dimension must be learned, which requires some notion of consistency---whether in a temporal (consecutive images in a video must maintain temporal consistency) or mechanical sense (stresses in consecutive deformation steps must ensure mechanical consistency). Analogous to a text descriptor prompting an image sequence, the nonlinear target response here serves as input to predict a sequence of mechanically deformed microstructural configurations along the deformation path, ultimately resulting in the effective stress-strain response. This requires the definition of an efficient design/property space to be considered as training data for our generative model, whose key concepts and the considered model architecture are summarized in the following.

\section{Results}
\label{sec:results}

\subsection{Generation of metamaterials with diverse properties}
\label{sec:metamaterial_generation}

As our diffusion framework operates in a data-driven setting, we require a large collection of paired mechanical designs and their corresponding nonlinear stress-strain responses. The options for potential design spaces are virtually unlimited, ranging from truss descriptors\cite{Bastek2022} over shells\cite{Kumar2020} to composite structures.\cite{Abueidda2019} We here consider a pixel-based design space parameterization with minimal constraints (aside from a periodic structure) to fully harness the generative power of diffusion models. While two-material composites could be generated with randomly drawn binary pixels and span a tremendous design space\cite{Abueidda2019}, the subset of structures with a non-trivial stress-strain response is comparably small. We therefore consider cellular structures (each pixel representing solid or void) as our design space to enable interesting mechanical behavior such as \textit{buckling}---an instability that quickly transitions between distinct equilibrium configurations---and \textit{contact}---arising under compressive loads and producing a sudden stiffness increase---overall resulting in a rich and possibly non-monotonic stress-strain curve. While modeling those effects by the FE method is challenging, inversely designing such structures is even more difficult due to the sensitivity of, e.g., the buckling response to small changes in the design. At the same time, incorporating such effects guarantees a highly diverse range of achievable stress-strain responses. To keep the problem tractable yet without loss of generality, we restrict our study to two dimensions (2D) and a periodic structure based on a square unit cell (UC).

\begin{figure}[h]
    \centering
    \includegraphics[width=1.0\textwidth]{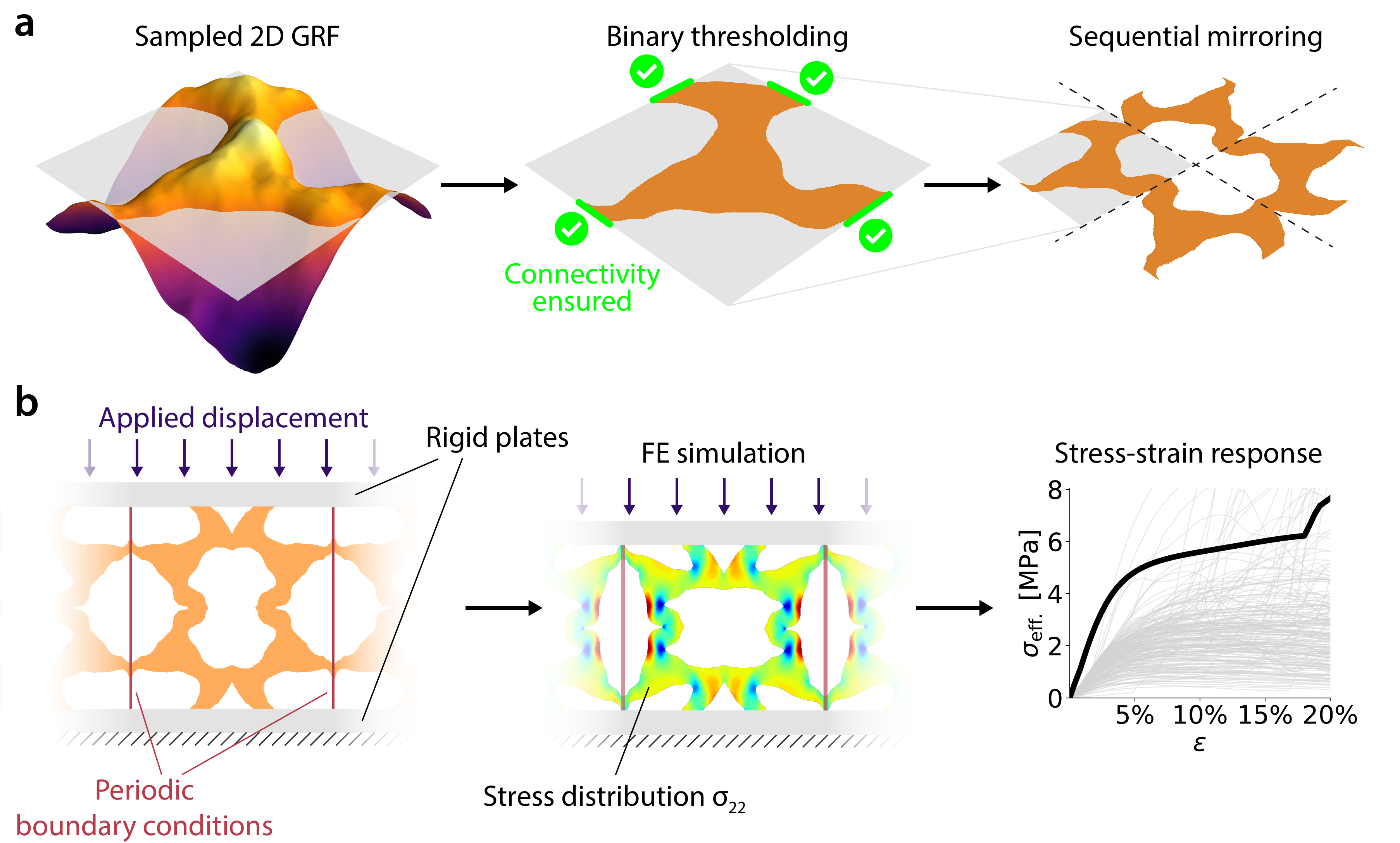}\caption{\textbf{Metamaterial generation process.} \textbf{(a)} A 2D cellular UC is generated by sampling from a 2D Gaussian random field, applying a varying threshold to extract a binary field, and mirroring the resulting pattern when connectivity to the boundaries is ensured. \textbf{(b)} To obtain the stress-strain response, we place the UC between two rigid plates with periodic boundary conditions in the horizontal direction and apply a compressive strain of up to $20\%$. The corresponding stress and displacement fields within the UC are computed by FE simulations, and the overall effective stress-strain response $\sigma_{\text{eff.}}$ is extracted from the nodal reaction forces, though they can be equally obtained from the full-field data. A representative selection of responses of the generated designs is plotted in gray.}
    \label{fig:metamaterial_generation}
\end{figure}

The generation of the dataset used for training is performed as follows (see Fig.~\ref{fig:metamaterial_generation}). To generate a random design with a certain level of structural features, we sample from a 2D Gaussian random field on a square domain and apply a binary threshold. Values above a specific threshold are considered material, those below are void. We ensure that opposite boundaries of the domain are connected with each other (and repeat the sampling until this condition is met) and mirror the pattern sequentially along both edges (see Fig.~\ref{fig:metamaterial_generation}) to obtain mechanically intricate, periodic structures. Despite its simplicity, this stochastic approach produces a diverse dataset of designs with a broad range of stress-strain responses. We further induce different levels of relative density (or fill fraction) by randomly shifting the threshold within a specified range. Higher values promote low-density structures prone to buckling, which is important for the aforementioned reasons.

The stress-strain response of each design is obtained from FE simulations. As a technologically relevant load case, we place all samples between two rigid plates and apply a quasistatic compressive strain of up to $\varepsilon=20\%$ in the vertical direction. Uniaxial compression is a frequent load characteristic of, e.g., impact applications\cite{Wang2022}, the compression of shoe soles\cite{Xiao2022}, or so-called \textit{passive compliance} in soft robotics (e.g., allowing a soft gripper to adapt its shape to the object being grabbed\cite{Liu2020}). By applying periodic boundary conditions along the horizontal directions, we simulate an infinite periodic layer of the chosen design, as found in sandwich-type configurations. Within the cellular UC, we account for frictional contact and use an experimentally calibrated elastoplastic material model \cite{Jin2020} (representative of a thermoplastic resin) to ensure realistic responses. Simulation details are provided in Methods.

Using this setup, we generate 53{,}007 pairs of unique designs and the corresponding stress-strain responses. We also collect the full-field stress distribution in the vertical direction, $\sigma_{22}$, as well as displacement components $u_1$ and $u_2$ (all in the Lagrangian frame), since this data contains valuable information about the underlying physics, as also observed by Nie et al.\cite{Nie2020} The overall effective stress response can be extracted either from the nodal reaction forces or directly from the full-field data, since in the considered quasi-static setting internal forces must be in equilibrium for any free cut of the UC (e.g., for any pixel row, see Supplementary Information Section~S5.1). We evaluate all fields on a $96\times96$ pixel grid together with the overall (average) vertical stress at eleven equidistant strain increments between 0\footnote{Since there are no stresses at zero strain, we instead collect the fields at $\varepsilon=0.2\%$.} and 20\%. This strikes a reasonable balance between accuracy and computational feasibility and provides the training data for the generative model.

\subsection{Video denoising diffusion model}
\label{sec:denoising_diffusion}

\begin{figure}[h]
    \centering
    \includegraphics[width=1.0\textwidth]{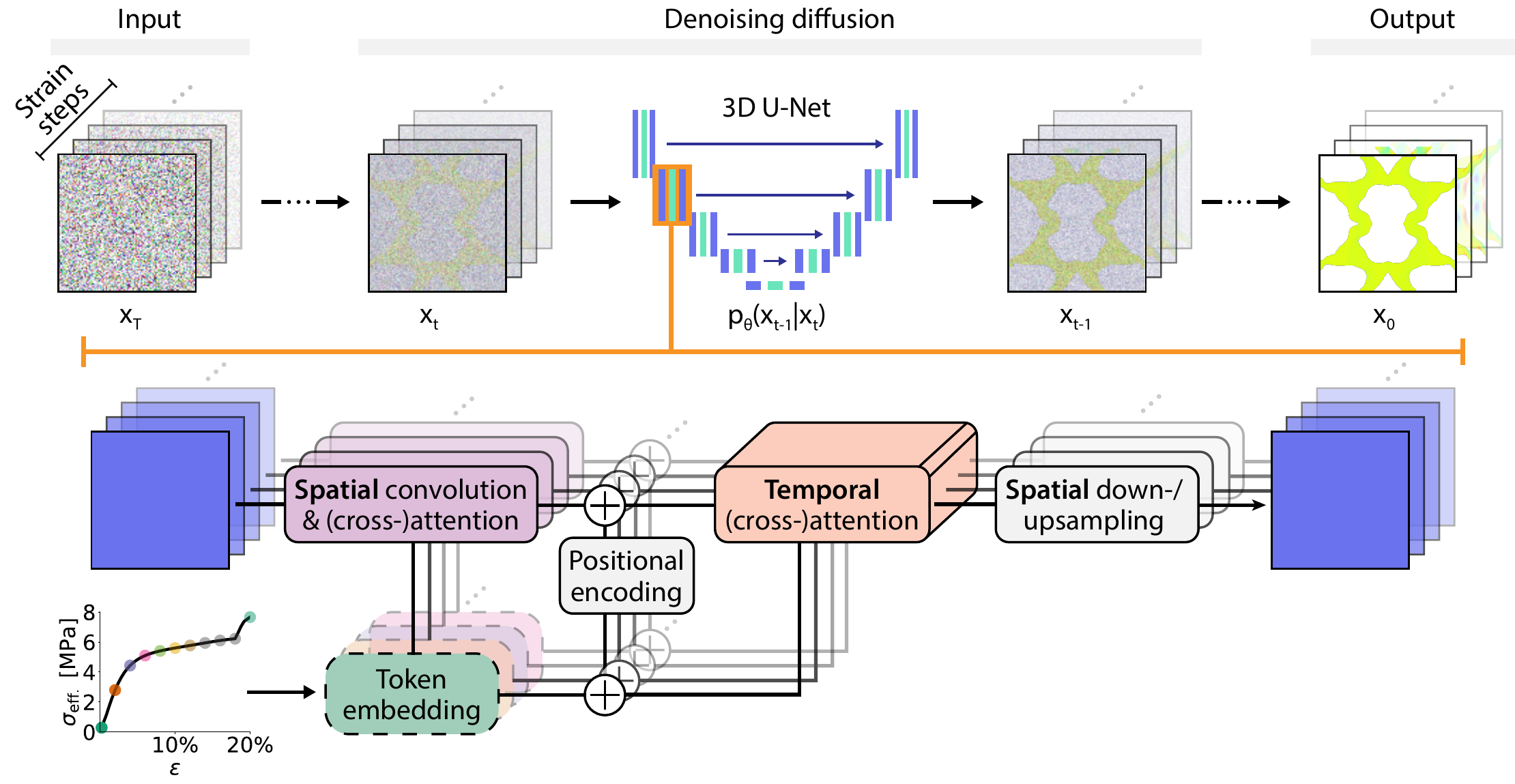}
    \caption{\textbf{Denoising diffusion model architecture.} The denoising diffusion model is based on the 3D U-Net video architecture\cite{Ho2022} which iteratively adds information to a Gaussian prior. To include a temporal dimension, each spatial convolution and attention layer is followed by temporal attention computed over the eleven strain steps. We condition the model by transforming the stress-strain response to a token embedding, which is added via cross-attention into both spatial and temporal attention layers.}
    \label{fig:denoising_diffusion_architecture}
\end{figure}

Diffusion models are trained to reverse a stochastic forward process that gradually converts a data point (i.e., an image) drawn from the underlying data distribution $\bfx_{0} \sim q(\bfx)$ to a prior distribution in $T$ steps, typically a standard Gaussian.\cite{Sohl-Dickstein2015,Ho2020} This can formally be understood as a fixed Markov chain with given variance schedule $\left\{\beta_t \in(0,1)\right\}_{t=1}^T$ as
\be
q(\bfx_{1:T} \vert \bfx_{0})=\prod_{t=1}^{T} q(\bfx_{t} \vert \bfx_{t-1}), \quad q(\bfx_{t} \vert \bfx_{t-1})=\mathcal{N}(\bfx_{t} ; \sqrt{1-\beta_{t}} \bfx_{t-1}, \beta_{t} \bfI).
\ee
This allows to sample $\bfx_{t}$ at any timestep $t$ via $\bfx_{t} =\sqrt{\bar{\alpha}_{t}} \bfx_{0}+\sqrt{1-\bar{\alpha}_{t}} \boldsymbol{\epsilon}$ with $\boldsymbol{\epsilon} \sim \mathcal{N}(\bf0, \bfI)$ and where $\alpha_{t}=1-\beta_{t}, \, \bar{\alpha}_{t}=\prod_{i=1}^{t} \alpha_{i}$.

We approximate the reverse process $q(\bfx_{t-1} \vert \bfx_{t})$ by a neural network $p_{\theta}(\bfx_{t-1} \vert \bfx_{t})$ parameterized by $\theta$. To generate new samples $\bfx^* \sim q(\bfx)$, we run the reverse Markov chain to arrive at
\be
p_{\theta}(\bfx_{0: T})=p(\bfx_{T}) \prod_{t=1}^{T} p_{\theta}(\bfx_{t-1} \vert \bfx_{t}), \quad p_{\theta}(\bfx_{t-1} \vert \bfx_{t})=\mathcal{N}(\bfx_{t-1} ; \boldsymbol{\mu}_{\theta}(\bfx_{t}, t), \boldsymbol{\Sigma}(\bfx_{t},t)).
\ee
Such models are typically trained to maximize the variational lower bound of the log-likelihood, which can be computed in closed form when conditioned on $\bfx_0$. As observed by Ho et al.\cite{Ho2020}, $\boldsymbol{\mu}_{\theta}$ can be decoupled into two terms relating to $\bfx_t$ and $\boldsymbol{\epsilon}_{\theta}$, allowing to simplify and re-parameterize the loss in terms of the Gaussian noise as
\be
\calL(\theta)=\mathbb{E}_{t, \bfx_{0}, \epsilon}\left[\left\lvert\boldsymbol{\epsilon}-\boldsymbol{\epsilon}_\theta\left(\bfx_t, t\right)\right\rvert\right].
\ee
To condition the model on some additional input $\bfc$, we consider \textit{classifier-free guidance}\cite{Ho2021}, not requiring an additional classifier $p_{\theta}(\bfc\vert\bfx_{t})$. We steer the reverse diffusion process by replacing $\boldsymbol{\epsilon}_{\theta}$ by a linear combination of the conditional and unconditional noise estimates, i.e.,
\be
\tilde{\boldsymbol{\epsilon}}_\theta\left(\bfx_t, \bfc\right)=(1+w) \boldsymbol{\epsilon}_\theta\left(\bfx_t, \bfc\right)-w \boldsymbol{\epsilon}_\theta\left(\bfx_t,\bfc=\emptyset)\right),
\ee
where $w\geq0$ is the guidance weight, allowing to trade-off sample quality with conditioning augmentation, and $\emptyset$ denotes a fixed random embedding to represent the lack of conditioning. Details are provided in Supplementary Information Section~S2.

Diffusion models map noisy input data to a less distorted one, making symmetric U-Net architectures\cite{Ronneberger2015} a common choice for $\boldsymbol{\epsilon}_\theta$. As our primary interest is in mapping from a target stress-strain curve to a \textit{design}, training the model on simple images of UCs conditioned on the corresponding stress-strain curve is a straightforward approach and has been explored in recent work.\cite{Vlassis2023} In our investigations, we observed similar success of such approaches for generating structures with a relatively simple stress-strain response (as the ones shown in \cite{Vlassis2023}). However, the same setup proved ineffective in modeling more challenging responses such as, e.g., those induced by contact and buckling. We attribute this limitation to the highly indirect mapping the model must learn to predict the full deformation history and the corresponding internal stress distributions, which in turn dictate the overall stress-strain response. To facilitate the training, to improve the sample efficiency, and to obtain a full-field prediction of the expected deformation \textit{path} and internal stresses for physical validation, we train the model not only on the UC design but also on the full-field data of the vertical stresses $\sigma_{22}$ for each strain step, as described in Section~\ref{sec:metamaterial_generation}. We observed best results when using a Lagrangian frame instead of a Eulerian one (i.e., evaluating all evolving fields on the undeformed initial configuration), which we additionally supply with the horizontal and vertical displacements $u_1$ and $u_2$. This allows us to optionally convert data to the Eulerian frame and provide information about the deformation path to the model.

Instead of simply concatenating this data along the image channels of the U-Net, we distinguish between the two fundamentally different causal relations of the data---space and applied strain---similar to recently proposed video generative models.\cite{Ho2022} Here, variants of the 2D (space-only) U-Net architecture are extended by a \textit{temporal} dimension, which effectively is treated as a batch axis and thus leaves the base architecture unaffected. The extension is a temporal attention\cite{Vaswani2017} block (taking the pixels as batch axis and computing self-attention over the applied strain steps) after the spatial attention and convolution (taking the strain steps as batch axis and computing convolutions and self-attention over the pixels) to learn physical consistency across different strain steps.

This architecture allows for mechanically motivated conditioning of the model on a given nonlinear stress-strain response. The conditioned effective stress at the eleven strain steps is directly associated with the corresponding full-field response\footnote{Mechanical equilibrium requires that the effective, overall stress at any strain level matches the averages of all pixel stress values across any row of pixels in the UC.}, which we leverage in our model architecture (unlike in video generation, in which words, as conditioning, do not directly correspond to specific image frames). To do so, we convert each stress value to a high-dimensional token embedding by a (learnable) linear layer and fuse it with the pixel representation via cross-attention\cite{Vaswani2017} in the spatial attention module of the corresponding strain step. In the subsequent temporal attention layer across all strain steps, we add a relative position encoding\cite{Shaw2018} to both the strain steps and token embeddings, so that the model receives information on the strain step order, and we apply ``pseudo-temporal'' cross-attention over the strain steps. Lastly, we augment the conditioning by adding a latent representation of the tokens to the diffusion time embedding (required as input to the model to indicate the diffusion time step). For further details see Methods, Supplementary Information Section~S3 and S4, and Code Availability.

\subsection{Full-field predictions for generated metamaterials}
\label{sec:full_field_pred}

A key advantage of our setup over other deep-learning frameworks is its capability to provide physical insight into the deformation mechanisms of the generated metamaterial and the associated stress response. By reversing the diffusion process conditioned on the desired stress-strain curve, we obtain not only a potential design but also a predicted full-field $\sigma_{22}$-distribution subjected to the applied strain throughout the deformation path. This enables us to evaluate the proposed deformation mechanism for physical validity and extract the \textit{predicted} stress-strain response by row-wise pixel averaging of the internal stress $\sigma_{22}$. In contrast to alternative approaches\cite{Vlassis2023}, our framework unifies inverse design and forward prediction in a single model without the need for an ad-hoc secondary model to evaluate the performance of the predicted designs. This also allows for the adoption of further design criteria (e.g., enforcing a maximum local stress to prevent failure).

We demonstrate the ability of the model to predict designs matching a given target stress-strain response by considering 100 \textit{responses} of randomly generated designs (unseen during training). We plot four predictions in Fig.~S6 of Supplementary Information and compute the average \textit{Normalized Root Mean Square Error} (NRMSE, see Methods) of the FE-reconstructed response vs.\ the target response as $\epsilon=6.98\%$. This is close to the mismatch of $\epsilon=2.74\%$ between the predicted and target responses, which underlines the model's ability to propose designs and concurrently estimate their mechanical behavior. The agreement between the predicted and true (i.e., high-fidelity FE) responses suggests an accurate estimate of the stress distribution, confirmed both qualitatively in Fig.~S6 of Supplementary Information and quantitatively with $\epsilon_{\text{L}_{2}}=14.39\%$, averaged over all samples and strain steps. (Supplementary Information Section~S6.1 summarizes a similar study on unconditionally sampled designs.)

\subsection{Inverse design of unseen stress-strain responses}
\label{sec:property_synthesis}

%\begin{figure}[ht!]
\begin{figure}
    \centering
    \includegraphics[width=1.0\textwidth]{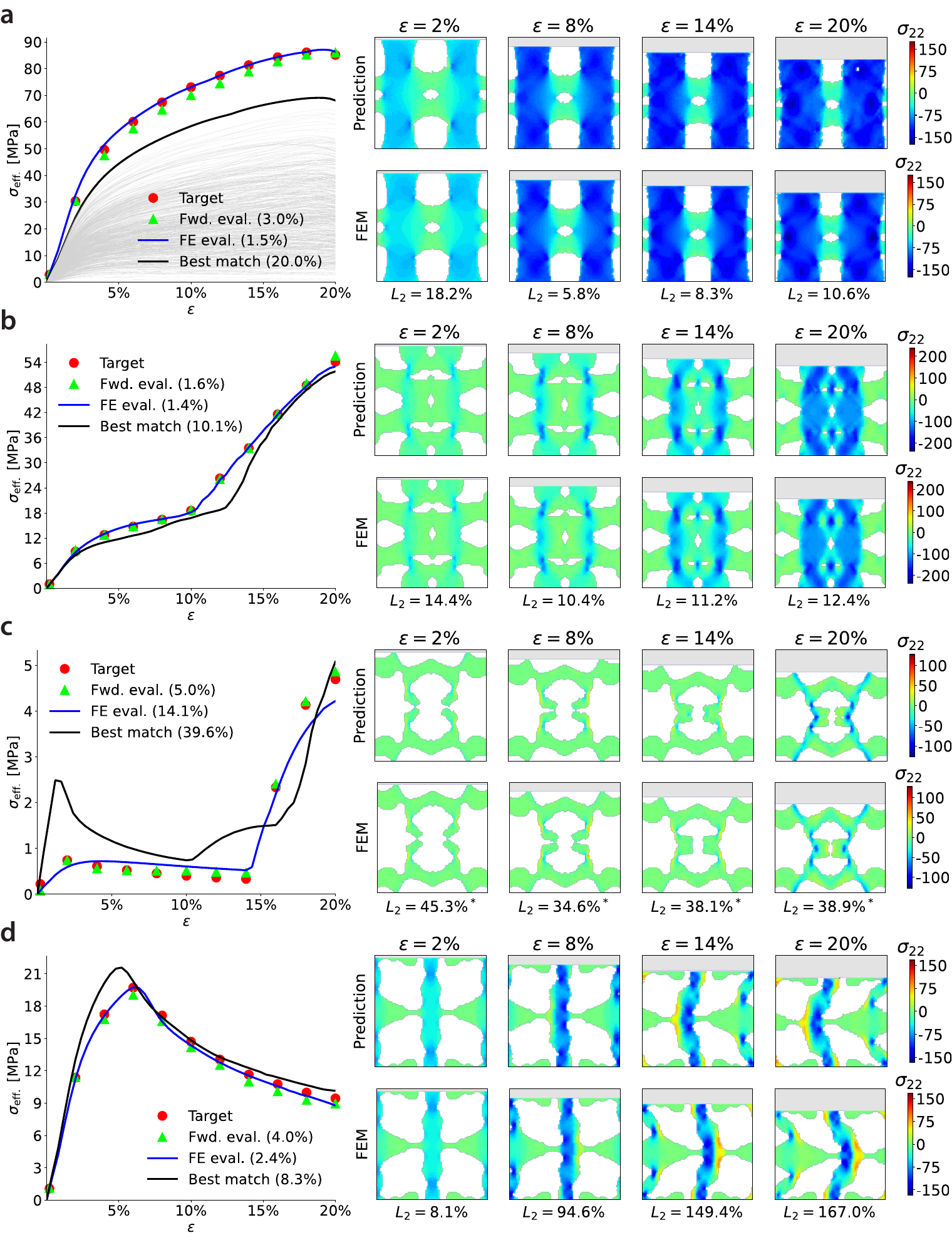}
    \caption{\textbf{Metamaterial synthesis for four stress-strain responses not represented in the training dataset.} \mbox{\textbf{(a-d)}} The model is conditioned on four technically relevant, challenging target responses. Validation of the predicted effective stress response $\sigma_{\text{eff.}}$ (`Fwd. eval.'; NRMSE with respect to the target response in brackets) of the generated designs is achieved by FE simulations (`FE eval.'), agreeing with the predicted response and significantly outperforming the best match in the training dataset (`Best match'). We additionally compare the predicted full-field $\sigma_{22}$-distribution (indicated in MPa in the Eulerian frame) with the FE ground truth and provide the corresponding relative $L_2$-errors. To highlight the range of responses in the training dataset, we plot a representative selection in gray in \textbf{(a)}. $^*$The relative $L_2$-error is numerically inflated due to the small magnitude of the stress field and is hence not truly indicative (but included for completeness).}
    \label{fig:diffusion_evaluation}
\end{figure}

The above results provide only a limited measure of the model's generalization performance: although the conditioned stress-strain responses are based on designs not seen during training, they are, on average, well-represented by samples in the training data. To assess the model's generalization capability, we next examine its performance on such responses not closely represented in the training data. We create four benchmark examples of diverse stress-strain responses that cover a wide range of material responses of engineering interest and include the non-trivial mechanisms of contact and buckling. For each case, we leverage the probabilistic nature of the model and generate ten samples conditioned on the target response and plot the best match. A guidance weight of $w=5$ was observed to enhance the match between generated design and target response without sacrificing the accuracy of the generated full-field predictions.

First, we generate a design with high stiffness, strong (nonlinear) hardening, and large deformability, as used, e.g., in impact applications. We condition the model with an effective stress response 20\% above the stiffest sample of the training set. As illustrated in Fig.~\ref{fig:diffusion_evaluation}a, the model generates a structure with a large fill fraction, closely matching the ground truth in both the FE-reconstructed response (with $\epsilon=1.5\%$; compared to $\epsilon=20\%$ of the best match in the training data) and the underlying stress distribution (ranging from $\epsilon_{\text{L}_2}=18.2\%$ to $\epsilon_{\text{L}_2}=5.8\%$). Analogously, compliant low-density designs can be generated by choosing a target stress-strain response well below the most compliant design in the training data (see Supplementary Information Section~S6.4), which is matched with $\epsilon=4.3\%$.

Second, we consider a more complex target response exhibiting an abrupt stiffness increase midway through the loading path (at 10\% applied strain, see Fig.~\ref{fig:diffusion_evaluation}b), which necessitates a change in deformation mode. Such stiffness changes can be leveraged, e.g., in soft robotic grippers.\cite{Firouzeh2017} The design proposed by the model indeed closely matches the target response ($\epsilon=1.4\%$) and significantly outperforms the closest match in the training data ($\epsilon=10.1\%$). Moreover, we observe that the generated design contains a fillet in its interior, which establishes contact at 10\% strain in both forward prediction and FE simulation, leading to the desired stiffness increase. This demonstrates that the model can introduce new contact mechanisms to match unseen responses, while---importantly---contact has so far been outside the scope of, e.g., computational TO.

Third, we consider the more exotic target of a highly compliant response until 15\% strain, followed by a drastic stiffness increase. (Such behavior can be caused by contact within the UC but is also characteristic of, e.g., structural transformations in metals.\cite{Hu2023}) While, as expected, the generated design is not as close as the previous targets ($\epsilon=14.1\%$), it considerably outperforms the best match in the training set ($\epsilon=39.6\%$). The initial compliance and sudden stiffness increase are realized through a delicate interplay of an almost purely rotational, auxetic response of an inner segment of the UC and the subsequent emergence of contact at the critical strain level where hardening sets in (see Fig.~\ref{fig:diffusion_evaluation}c).

Fourth, we consider a response with significant \textit{softening}, which is utilized, e.g., in snapping and release mechanisms. As illustrated in Fig.~\ref{fig:diffusion_evaluation}d, the model's design again outperforms the best match ($\epsilon=2.4\%$ vs.\ $\epsilon=8.3\%$). The response is accommodated by a buckling mechanism. Interestingly, the relative $L_2$-error of the predicted stress fields significantly increases in the post-buckling regime. This, however, stems from the symmetric buckling mode of the design and the fact that the FE simulation buckles to the right while the model predicts buckling to the left (Buckling is highly sensitive to the design (unlike contact): when a vertical column is compressed in 2D, it can buckle to the left or to the right, sensitive to smallest imperfections.). In this case, we cannot reasonably expect the model to match this response. Instead, this demonstrates its temporal consistency and logically completes the deformation trajectory---once buckled to the right, the post-buckling follows this trend. (An example of a generated design with a predicted deformation mode matching the FE simulation is shown in Supplementary Information Section~S6.5.)

\section{Discussion}
\label{sec:discussion}

Soft robots and biomimetic structures, among others, require materials with precise nonlinear mechanical functionality---a challenge for conventional optimization techniques due to the complex inherent deformation mechanics including buckling and contact. Gradient-based optimizers may become numerically unstable due to the nonlinear and non-convex objective function. This issue worsens when considering contact, which leads to abrupt, non-smooth kinks in the stress response. Our model, inspired by generative video modeling, is particularly suited to this nonlinear setting. It accurately captures the non-trivial mechanics at play and unifies an efficient surrogate forward model with the ability to generate unseen metamaterial designs exhibiting complex nonlinear responses, which must leverage buckling and contact. This is accomplished by training the model on the complete deformation trajectory rather than solely on the underlying designs (akin to extending image to video generative models), which may suffice for linear conditioning but is inadequate for complex nonlinear situations (see the ablation study in Supplementary Information Section~S7).

The complex target responses may be associated with multiple designs, posing a challenge for direct optimization. Addressing this one-to-many mapping is a recurring issue in inverse problems across disciplines, for which the probabilistic nature inherent in the diffusion architecture is ideally suited. By repeatedly generating samples for identical target responses, our model proposes a variety of designs (which may be checked for secondary objectives such as manufacturability). Our work further demonstrates the efficacy of video diffusion models when data of different modalities, such as the effective stress-strain response and the full-field internal stress distribution, must be synthesized and optimized---a task where conventional optimization techniques may fail. 

The presented framework admits extension to related fields such as fluid dynamics, serving both as a surrogate simulator and nonlinear optimizer. The efficiency of the denoising diffusion process may be increased by operating in a latent space, which may include additional data modalities in the conditioning such as the underlying base material. Moreover, alternative design spaces such as trusses\cite{Bastek2022} provide a more compact design parameterization for 3D structures and low fill fractions. As trusses can naturally be represented by graphs, graph diffusion models, mainly used in molecule design, can serve as a viable model architecture.

\section{Methods}
\label{sec:methods}

We here provide details of the data generation procedure, the methods employed for creating the metamaterials under consideration, and the FE setup to evaluate the nonlinear mechanical response of UCs. We further present the model architecture as well as the training and sampling protocol. Additional explanations can be found in Supplementary Information.

\subsection{Design generation}
\label{sec:design_generation}

We generate a random mechanical metamaterial by sampling a 2D Gaussian Random Field (GRF) on a square domain based on the algorithm proposed by Lang \& Potthoff.\cite{Lang2011} To do so, we sample complex Gaussian noise for a centered (even) $N \times N$ grid of Fourier coordinates
\begin{equation*}
\calK = \left\{ (k_1, k_2) \in \mathbb{Z}^2 \, : \, -N/2 \leq k_1 < N/2, -N/2 \leq k_2 < N/2 \right\}
\end{equation*}
and introduce spatial correlation by a power law of the type $P(k_1, k_2) \propto (k_1^2+k^2_2)^{-\alpha/2}$, where we set $\alpha=3$ to ensure sufficient smoothness for manufacturable structures. This representation is converted to the corresponding real $N \times N$ pixel set $\calX$ by considering the standardized real part of the inverse Discrete Fourier Transform (DFT). Next, we convert it to binary values ($1$~representing material, $0$~representing void) by considering a threshold $t$ sampled as $t \sim \mathcal{U}(0,t_{\text{max}})$ with $t_{\text{max}}=3/5$, which was chosen to increase the variance (in terms of sparsity) of the sampled structures. Lastly, we check for the connectedness of the four boundaries of the square grid, which is defined as given if there exists a single material domain that covers at least $10\%$ of the pixels (rounded down) of each side. This avoids structures with extremely sparse connectivity (and hence questionable manufacturability). We repeat the process until a valid structure has been found. The metamaterial is created by mirroring the found structure sequentially along the vertical and horizontal boundaries to ensure periodicity. While we only focus on periodicity in the horizontal direction in the examples presented in this work, the generated structures can also be tessellated along the vertical direction to produce 2D tessellations. Note that the GRFs are by construction periodic, so they can also be tessellated without mirroring. However, we found that mirroring generates in general more diverse stress-strain responses and further simplifies the mesh generation for periodic boundary conditions, which is why we chose this procedure. The pseudocode of this process is given in Algorithm~1 in Supplementary Information Section~S1.

\subsection{FE simulations}
\label{sec:fe_simulation}

To evaluate the stress-strain responses of the generated structures, we use Abaqus CAE 2020. All of the following steps are implemented via User Subroutines. Note that we apply a smoothening of the boundary of the generated pixel structures to bypass issues with the meshing, presented in Supplementary Information Section~S1.2. We generate a mesh compatible with periodic boundary conditions (i.e., featuring matching nodes on opposite boundaries) and select 3-node linear (\texttt{CPE3}) and 4-node bilinear elements with reduced integration and hourglass control (\texttt{CPE4R}) using default settings. The mesh was refined until sufficient convergence in the stress distributions and overall stress-strain responses was observed. We consider plane-strain conditions to represent the realistic scenario of an extruded structure in the out-of-plane dimension (thus avoiding challenges with out-of-plane buckling under compression).

The metamaterial is virtually positioned between two rigid horizontal platens, to which we attach the nodes on the top and bottom boundary. We assume lubricated surfaces, so that nodes may slip horizontally relative to the horizontal platens. Within the UC, we consider frictional self-contact with a friction coefficient $k_\text{fric.}= 0.4$. Due to the presence of large deformations including buckling and contact, an implicit dynamic solver is chosen for numerical stability. We ensure a quasi-static simulation by setting the mass density to $\rho=10^{-8}$, applying displacements with a smooth amplitude from time $t=0$ to $t=1$, and confirming that the kinetic energy (\texttt{ALLKE}) does not exceed $1\%$ of the internal energy (\texttt{ALLIE}) for all strain steps.
We furthermore verify that artificial energy measures (\texttt{ALLAE} and \texttt{ALLSD}), introduced for stability reasons, do not individually exceed $1\%$ of the internal energy across all strain steps. In general, we use unit-less values for all lengths in simulations (due to size invariance), while stresses are presented in units of MPa.

We record the horizontal and vertical displacement components ($u_{1}$ and $u_{2}$, respectively), as well as the vertical stress component $\sigma_{22}$ on a $96\times96$ pixel grid at eleven evenly placed strain increments from the undeformed configuration to the total applied vertical strain in the Lagrangian (undeformed reference) frame. Note that instead of taking the initial step at 0\% strain, we consider all fields at $0.2\%$ strain, as this provides information on the small-strain response of the structure instead of trivial all-zero values. To compute the effective, overall stress response (which is the net vertical force per initial (undeformed) area on the top or bottom surfaces) at any strain level, we record the vertical reaction forces (\texttt{RF2}) of those nodes in contact with the upper rigid surface. Details on the considered base material can be found in Supplementary Information Section~S1.3. All simulations were carried out on the Euler high-performance cluster of ETH Zurich.

\subsection{Spatial 2D U-Net architecture}
We refer to the Code Availability Section for full technical details and below provide a high-level summary of the denoising diffusion model architecture. The PyTorch framework\cite{Paszke2019} was used throughout our implementation. Diffusion models iteratively remove noise from data, typically images. Consequently, their input and output dimensions must be equal, making U-Net architectures a prevalent choice. Our model builds upon the work of Ho et al.\cite{Ho2022} and its implementation provided by Phil Wang\cite{Wang2022a}, which, in turn, are based on derivations of the original 2D-U-Net architecture\cite{Ronneberger2015}. This encoder-decoder architecture incrementally reduces spatial information while increasing latent feature information before reversing this operation by reducing the latent representation back to the spatial domain. In our work, each down- and upsampling pass comprises two ResNet\cite{Zagoruyko2016} blocks consisting of a series of convolutional layers and SiLU activation functions\cite{Elfwing2017}, spatial linear self-attention\cite{Katharopoulos2020} (to reduce computational complexity) across the (latent) pixel representation, and a down- or upsampling convolutional layer. The middle block between the encoder and decoder equally consists of two Resnet blocks with a (full) spatial self-attention layer in-between. We use four feature map resolutions ($96\times96 \rightarrow 12\times12$) with expanding latent dimensions $(64 \rightarrow 512)$. Each attention block consists of 8 attention heads, each with a dimension of 32. We summarize the most relevant hyperparameters in Table~S2 in Supplementary Information Section~S3.

\subsection{Extension to temporal 3D U-Net architecture}
We extend the 2D U-Net by incorporating a temporal dimension\cite{Ho2022}, where we understand the ``temporal'' dimension as the applied strain steps. In all building blocks described above, the temporal dimension is treated as a batch dimension and therefore does not affect the setup. The key difference is that we insert a temporal self-attention layer at the beginning before the encoder-decoder architecture and additionally, after every spatial attention layer, which treats the spatial dimension as batch axes and performs attention over the eleven strain steps. We consider relative positional encoding\cite{Shaw2018} to pass information of the strain step order to the model.

\subsection{Conditioning on nonlinear stress-strain responses}
To condition the model on the stress-strain response, we convert all eleven scalar stress values at the corresponding strain steps to an embedding via a (learnable) linear layer. Note that we omit the corresponding strain value since we keep these fixed in this work, thus providing no further information, though a future extension can explore adaptive stepping techniques, such as sampling more densely at strain steps with significant deformation changes. These token embeddings are concatenated to the spatial attention tokens at the corresponding strain step for cross-attention, while we concatenate all eleven token embeddings with a relative positional encoding to the temporal attention tokens in the temporal attention layer. Note that for cross-attention we derive the queries from the pixel embedding but the keys and values from the conditioning embedding.
To further enhance the conditioning, we average all eleven token embeddings over the strain steps and convert this to a latent representation by a two-layer MLP and SiLU activation function\cite{Elfwing2017}, which transforms this representation to the same dimension as the latent embedding of the diffusion time step $t$. The latter is necessary for the model to determine the current step of the denoising process. We add both embeddings and incorporate them into the ResNet blocks.

\subsection{Training protocol}
\label{sec:training_prot}

We first pre-process the data as follows. We apply a min-max normalization to transform all input data $\bfx$ (i.e., stress and displacement distributions) and conditioning (i.e., stress-strain responses) to the range $[-1,1]$, i.e., 
\be \label{min_max_2}
   x_i \leftarrow \frac{2\left[x_i -\min(\boldsymbol{x})\right]}{\max(\boldsymbol{x})-\min(\boldsymbol{x})}-1,
\ee
where the min and max operators are applied across all corresponding data points. For the stress and displacement fields, we consider all corresponding pixel values for all strain steps in the entire training dataset. For the stress-strain responses, we consider the minimum and maximum recorded stress response for all strain steps in the entire training dataset. Note that we store the image/video data generated with Abaqus in the \texttt{gif} format to reduce storage requirements.

We provide the training hyperparameters in Table~S3 and the loss plots in Supplementary Information Section~S4.
The model was trained on the Euler high-performance cluster of ETH Zurich, utilizing parallel and mixed precision processing. We use the \texttt{Accelerate} library from Hugging Face to facilitate the training setup, which was conducted on eight Nvidia Quadro RTX 6000 GPUs, each equipped with 24 GB GDDR6 memory. The training process took approx.\ 70h.

\subsection{Sampling protocol}

Since the model does not directly predict binary pixels but stress and displacement distributions (which may be close to zero at the initial deformation stages), we require a robust method of extracting the underlying (undeformed) structure. We achieve this by considering the vertical displacement $u_{2}$ of the upper left quarter (corresponding to the gray area in Fig.~1a) of the predicted field, which is sufficient to extract the full topology due to symmetry. For each pixel, we check whether its value is within a $2\%$ tolerance around zero displacement (relative to the maximum displacement range) across all strain steps. If so, we consider it void (and otherwise material). We found this method to be highly robust, as the upper boundary of the structure is compressed and thus all `material pixels' will likely undergo some level of displacement (exceeding the set tolerance). We remove any disconnected sub-domains of the obtained design (though these were rarely observed). Further details on the effective stress response prediction and the mitigation of accuracy losses are provided in Supplementary Information Section~S5.

\subsection{Error measures}

To obtain an objective and scale-invariant error norm of the stress-strain curves, we consider the \textit{Normalized Root Mean Square Error} (NRMSE) computed as
\be
\epsilon(\boldsymbol{\sigma}^{\text{pred.}}_{\text{eff.}},\boldsymbol{\sigma}^{\text{true}}_{\text{eff.}}) = \sqrt{\frac{\lVert \boldsymbol{\sigma}^{\text{pred.}}_{\text{eff.}}-\boldsymbol{\sigma}^{\text{true}}_{\text{eff.}}\lVert^2}{\lVert\boldsymbol{\sigma}^{\text{true}}_{\text{eff.}}\lVert^2}},
\ee
where $\boldsymbol{\sigma}_{\text{eff.}} \in \mathbb{R}^{11}$ is the vector collecting the effective stress values $\sigma_{\text{eff.}}$ at the eleven strain steps, and $\lVert\cdot\lVert$ is the Euclidean norm.

For the full-field responses, we compute the analogous relative $L_2$-error per strain step as
\be\epsilon_{L_{2}}(\boldsymbol{\sigma}^{\text{pred.}}_{22},\boldsymbol{\sigma}^{\text{true}}_{22}) = \sqrt{\frac{\lVert \boldsymbol{\sigma}^{\text{pred.}}_{22}-\boldsymbol{\sigma}^{\text{true}}_{22}\lVert^2}{\lVert\boldsymbol{\sigma}^{\text{true}}_{22}\lVert^2}},
\ee
where $\boldsymbol{\sigma}_{22} \in \mathbb{R}^{N\times N}$ denotes the $\sigma_{22}$-stress values of the discretized pixel grid in the Lagrangian frame for the corresponding strain step, and $\lVert\cdot\lVert$ is the Frobenius norm.

\section{Supplementary Information}
\label{sec:supp_information}
We will release the Supplementary Information along with the publication.

\section{Data availability}
\label{sec:data_availability}
We will release the training and validation dataset consisting of pairs of full-field data and the effective stress-strain response along with the publication.

\section{Code availability}
\label{sec:code_availability}
We provide the code used to train the model and generate new metamaterial designs conditioned on a given stress-strain response in \url{https://github.com/jhbastek/VideoMetamaterials}.

\begin{appendices}
%\section{Section title of first appendix}
%\label{secA1}




\end{appendices}


%\bibliography{bibliography.bib}

\begin{thebibliography}{50}
\ifx \bisbn   \undefined \def \bisbn  #1{ISBN #1}\fi
\ifx \binits  \undefined \def \binits#1{#1}\fi
\ifx \bauthor  \undefined \def \bauthor#1{#1}\fi
\ifx \batitle  \undefined \def \batitle#1{#1}\fi
\ifx \bjtitle  \undefined \def \bjtitle#1{#1}\fi
\ifx \bvolume  \undefined \def \bvolume#1{\textbf{#1}}\fi
\ifx \byear  \undefined \def \byear#1{#1}\fi
\ifx \bissue  \undefined \def \bissue#1{#1}\fi
\ifx \bfpage  \undefined \def \bfpage#1{#1}\fi
\ifx \blpage  \undefined \def \blpage #1{#1}\fi
\ifx \burl  \undefined \def \burl#1{\textsf{#1}}\fi
\ifx \doiurl  \undefined \def \doiurl#1{\url{https://doi.org/#1}}\fi
\ifx \betal  \undefined \def \betal{\textit{et al.}}\fi
\ifx \binstitute  \undefined \def \binstitute#1{#1}\fi
\ifx \binstitutionaled  \undefined \def \binstitutionaled#1{#1}\fi
\ifx \bctitle  \undefined \def \bctitle#1{#1}\fi
\ifx \beditor  \undefined \def \beditor#1{#1}\fi
\ifx \bpublisher  \undefined \def \bpublisher#1{#1}\fi
\ifx \bbtitle  \undefined \def \bbtitle#1{#1}\fi
\ifx \bedition  \undefined \def \bedition#1{#1}\fi
\ifx \bseriesno  \undefined \def \bseriesno#1{#1}\fi
\ifx \blocation  \undefined \def \blocation#1{#1}\fi
\ifx \bsertitle  \undefined \def \bsertitle#1{#1}\fi
\ifx \bsnm \undefined \def \bsnm#1{#1}\fi
\ifx \bsuffix \undefined \def \bsuffix#1{#1}\fi
\ifx \bparticle \undefined \def \bparticle#1{#1}\fi
\ifx \barticle \undefined \def \barticle#1{#1}\fi
\bibcommenthead
\ifx \bconfdate \undefined \def \bconfdate #1{#1}\fi
\ifx \botherref \undefined \def \botherref #1{#1}\fi
\ifx \url \undefined \def \url#1{\textsf{#1}}\fi
\ifx \bchapter \undefined \def \bchapter#1{#1}\fi
\ifx \bbook \undefined \def \bbook#1{#1}\fi
\ifx \bcomment \undefined \def \bcomment#1{#1}\fi
\ifx \oauthor \undefined \def \oauthor#1{#1}\fi
\ifx \citeauthoryear \undefined \def \citeauthoryear#1{#1}\fi
\ifx \endbibitem  \undefined \def \endbibitem {}\fi
\ifx \bconflocation  \undefined \def \bconflocation#1{#1}\fi
\ifx \arxivurl  \undefined \def \arxivurl#1{\textsf{#1}}\fi
\csname PreBibitemsHook\endcsname

%%% 1
\bibitem{Tancogne-Dejean2018}
\begin{barticle}
\bauthor{\bsnm{Tancogne-Dejean}, \binits{T.}},
\bauthor{\bsnm{Diamantopoulou}, \binits{M.}},
\bauthor{\bsnm{Gorji}, \binits{M.B.}},
\bauthor{\bsnm{Bonatti}, \binits{C.}},
\bauthor{\bsnm{Mohr}, \binits{D.}}:
\batitle{{3D Plate-Lattices: An Emerging Class of Low-Density Metamaterial
  Exhibiting Optimal Isotropic Stiffness}}.
\bjtitle{Advanced Materials}
\bvolume{30}(\bissue{45}),
\bfpage{1803334}
(\byear{2018}).
\doiurl{10.1002/adma.201803334}
\end{barticle}
\endbibitem

%%% 2
\bibitem{Kumar2020}
\begin{barticle}
\bauthor{\bsnm{Kumar}, \binits{S.}},
\bauthor{\bsnm{Tan}, \binits{S.}},
\bauthor{\bsnm{Zheng}, \binits{L.}},
\bauthor{\bsnm{Kochmann}, \binits{D.M.}}:
\batitle{{Inverse-designed spinodoid metamaterials}}.
\bjtitle{npj Computational Materials}
\bvolume{6}(\bissue{1}),
\bfpage{1}--\blpage{10}
(\byear{2020}).
\doiurl{10.1038/s41524-020-0341-6}
\end{barticle}
\endbibitem

%%% 3
\bibitem{Kadic2019}
\begin{barticle}
\bauthor{\bsnm{Kadic}, \binits{M.}},
\bauthor{\bsnm{Milton}, \binits{G.W.}},
\bauthor{\bparticle{van} \bsnm{Hecke}, \binits{M.}},
\bauthor{\bsnm{Wegener}, \binits{M.}}:
\batitle{{3D metamaterials}}.
\bjtitle{Nature Reviews Physics}
\bvolume{1}(\bissue{3}),
\bfpage{198}--\blpage{210}
(\byear{2019}).
\doiurl{10.1038/s42254-018-0018-y}
\end{barticle}
\endbibitem

%%% 4
\bibitem{Wu2021}
\begin{barticle}
\bauthor{\bsnm{Wu}, \binits{J.}},
\bauthor{\bsnm{Sigmund}, \binits{O.}},
\bauthor{\bsnm{Groen}, \binits{J.P.}}:
\batitle{{Topology optimization of multi-scale structures: a review}}.
\bjtitle{Structural and Multidisciplinary Optimization}
\bvolume{63}(\bissue{3}),
\bfpage{1455}--\blpage{1480}
(\byear{2021}).
\doiurl{10.1007/s00158-021-02881-8}
\end{barticle}
\endbibitem

%%% 5
\bibitem{Telgen2022}
\begin{botherref}
\oauthor{\bsnm{Telgen}, \binits{B.}},
\oauthor{\bsnm{Sigmund}, \binits{O.}},
\oauthor{\bsnm{Kochmann}, \binits{D.M.}}:
{Topology Optimization of Graded Truss Lattices Based on On-the-Fly
  Homogenization}.
Journal of Applied Mechanics
\textbf{89}(6)
(2022).
\doiurl{10.1115/1.4054186}
\end{botherref}
\endbibitem

%%% 6
\bibitem{DaSilva2022}
\begin{barticle}
\bauthor{\bparticle{da} \bsnm{Silva}, \binits{G.A.}},
\bauthor{\bsnm{Beck}, \binits{A.T.}},
\bauthor{\bsnm{Sigmund}, \binits{O.}}:
\batitle{{Structural topology optimization with predetermined breaking
  points}}.
\bjtitle{Computer Methods in Applied Mechanics and Engineering}
\bvolume{400},
\bfpage{115610}
(\byear{2022}).
\doiurl{10.1016/j.cma.2022.115610}
\end{barticle}
\endbibitem

%%% 7
\bibitem{Bastek2022}
\begin{barticle}
\bauthor{\bsnm{Bastek}, \binits{J.-H.}},
\bauthor{\bsnm{Kumar}, \binits{S.}},
\bauthor{\bsnm{Telgen}, \binits{B.}},
\bauthor{\bsnm{Glaesener}, \binits{R.N.}},
\bauthor{\bsnm{Kochmann}, \binits{D.M.}}:
\batitle{{Inverting the structure–property map of truss metamaterials by deep
  learning}}.
\bjtitle{Proceedings of the National Academy of Sciences}
\bvolume{119}(\bissue{1}),
\bfpage{2111505119}
(\byear{2022}).
\doiurl{10.1073/pnas.2111505119}
\end{barticle}
\endbibitem

%%% 8
\bibitem{Zheng2021}
\begin{barticle}
\bauthor{\bsnm{Zheng}, \binits{L.}},
\bauthor{\bsnm{Kumar}, \binits{S.}},
\bauthor{\bsnm{Kochmann}, \binits{D.M.}}:
\batitle{{Data-driven topology optimization of spinodoid metamaterials with
  seamlessly tunable anisotropy}}.
\bjtitle{Computer Methods in Applied Mechanics and Engineering}
\bvolume{383},
\bfpage{113894}
(\byear{2021}).
\doiurl{10.1016/j.cma.2021.113894}
\end{barticle}
\endbibitem

%%% 9
\bibitem{Tian2022}
\begin{barticle}
\bauthor{\bsnm{Tian}, \binits{J.}},
\bauthor{\bsnm{Tang}, \binits{K.}},
\bauthor{\bsnm{Chen}, \binits{X.}},
\bauthor{\bsnm{Wang}, \binits{X.}}:
\batitle{{Machine learning-based prediction and inverse design of 2D
  metamaterial structures with tunable deformation-dependent Poisson's ratio}}.
\bjtitle{Nanoscale}
\bvolume{14}(\bissue{35}),
\bfpage{12677}--\blpage{12691}
(\byear{2022}).
\doiurl{10.1039/D2NR02509D}
\end{barticle}
\endbibitem

%%% 10
\bibitem{Li2022}
\begin{botherref}
\oauthor{\bsnm{Li}, \binits{W.}},
\oauthor{\bsnm{Wang}, \binits{F.}},
\oauthor{\bsnm{Sigmund}, \binits{O.}},
\oauthor{\bsnm{Zhang}, \binits{X.S.}}:
{Digital synthesis of free-form multimaterial structures for realization of
  arbitrary programmed mechanical responses}.
Proceedings of the National Academy of Sciences
\textbf{119}(10)
(2022).
\doiurl{10.1073/pnas.2120563119}
\end{botherref}
\endbibitem

%%% 11
\bibitem{Elango2015}
\begin{barticle}
\bauthor{\bsnm{Elango}, \binits{N.}},
\bauthor{\bsnm{Faudzi}, \binits{A.A.M.}}:
\batitle{{A review article: investigations on soft materials for soft robot
  manipulations}}.
\bjtitle{The International Journal of Advanced Manufacturing Technology}
\bvolume{80}(\bissue{5-8}),
\bfpage{1027}--\blpage{1037}
(\byear{2015}).
\doiurl{10.1007/s00170-015-7085-3}
\end{barticle}
\endbibitem

%%% 12
\bibitem{Chan2008}
\begin{barticle}
\bauthor{\bsnm{Chan}, \binits{B.P.}},
\bauthor{\bsnm{Leong}, \binits{K.W.}}:
\batitle{{Scaffolding in tissue engineering: general approaches and
  tissue-specific considerations}}.
\bjtitle{European Spine Journal}
\bvolume{17}(\bissue{S4}),
\bfpage{467}--\blpage{479}
(\byear{2008}).
\doiurl{10.1007/s00586-008-0745-3}
\end{barticle}
\endbibitem

%%% 13
\bibitem{Zhang2020}
\begin{barticle}
\bauthor{\bsnm{Zhang}, \binits{J.}},
\bauthor{\bsnm{Lu}, \binits{G.}},
\bauthor{\bsnm{You}, \binits{Z.}}:
\batitle{{Large deformation and energy absorption of additively manufactured
  auxetic materials and structures: A review}}.
\bjtitle{Composites Part B: Engineering}
\bvolume{201},
\bfpage{108340}
(\byear{2020}).
\doiurl{10.1016/j.compositesb.2020.108340}
\end{barticle}
\endbibitem

%%% 14
\bibitem{Hyun-YongHan}
\begin{botherref}
\oauthor{\bsnm{{Hyun-Yong Han}}},
\oauthor{\bsnm{Kawamura}, \binits{S.}}:
{Analysis of stiffness of human fingertip and comparison with artificial
  fingers}.
In: IEEE SMC'99 Conference Proceedings. 1999 IEEE International Conference on
  Systems, Man, and Cybernetics (Cat. No.99CH37028),
vol. 2,
pp. 800--805.
IEEE.
\doiurl{10.1109/ICSMC.1999.825364}.
\url{http://ieeexplore.ieee.org/document/825364/}
\end{botherref}
\endbibitem

%%% 15
\bibitem{Yang2015}
\begin{barticle}
\bauthor{\bsnm{Yang}, \binits{D.}},
\bauthor{\bsnm{Mosadegh}, \binits{B.}},
\bauthor{\bsnm{Ainla}, \binits{A.}},
\bauthor{\bsnm{Lee}, \binits{B.}},
\bauthor{\bsnm{Khashai}, \binits{F.}},
\bauthor{\bsnm{Suo}, \binits{Z.}},
\bauthor{\bsnm{Bertoldi}, \binits{K.}},
\bauthor{\bsnm{Whitesides}, \binits{G.M.}}:
\batitle{{Buckling of Elastomeric Beams Enables Actuation of Soft Machines}}.
\bjtitle{Advanced Materials}
\bvolume{27}(\bissue{41}),
\bfpage{6323}--\blpage{6327}
(\byear{2015}).
\doiurl{10.1002/adma.201503188}
\end{barticle}
\endbibitem

%%% 16
\bibitem{Niu2019}
\begin{barticle}
\bauthor{\bsnm{Niu}, \binits{Z.}},
\bauthor{\bsnm{Wang}, \binits{X.}},
\bauthor{\bsnm{Meng}, \binits{X.}},
\bauthor{\bsnm{Guo}, \binits{X.}},
\bauthor{\bsnm{Jiang}, \binits{Y.}},
\bauthor{\bsnm{Xu}, \binits{Y.}},
\bauthor{\bsnm{Li}, \binits{Q.}},
\bauthor{\bsnm{Shen}, \binits{C.}}:
\batitle{{Controllable fiber orientation and nonlinear elasticity of
  electrospun nanofibrous small diameter tubular scaffolds for vascular tissue
  engineering}}.
\bjtitle{Biomedical Materials}
\bvolume{14}(\bissue{3}),
\bfpage{035006}
(\byear{2019}).
\doiurl{10.1088/1748-605X/ab07f1}
\end{barticle}
\endbibitem

%%% 17
\bibitem{Wang2014}
\begin{barticle}
\bauthor{\bsnm{Wang}, \binits{F.}},
\bauthor{\bsnm{Sigmund}, \binits{O.}},
\bauthor{\bsnm{Jensen}, \binits{J.S.}}:
\batitle{{Design of materials with prescribed nonlinear properties}}.
\bjtitle{Journal of the Mechanics and Physics of Solids}
\bvolume{69},
\bfpage{156}--\blpage{174}
(\byear{2014}).
\doiurl{10.1016/j.jmps.2014.05.003}
\end{barticle}
\endbibitem

%%% 18
\bibitem{Medina2023}
\begin{barticle}
\bauthor{\bsnm{Medina}, \binits{E.}},
\bauthor{\bsnm{Rycroft}, \binits{C.H.}},
\bauthor{\bsnm{Bertoldi}, \binits{K.}}:
\batitle{{Nonlinear shape optimization of flexible mechanical metamaterials}}.
\bjtitle{Extreme Mechanics Letters}
\bvolume{61},
\bfpage{102015}
(\byear{2023}).
\doiurl{10.1016/j.eml.2023.102015}
\end{barticle}
\endbibitem

%%% 19
\bibitem{Buhl2000}
\begin{barticle}
\bauthor{\bsnm{Buhl}, \binits{T.}},
\bauthor{\bsnm{Pedersen}, \binits{C.B.W.}},
\bauthor{\bsnm{Sigmund}, \binits{O.}}:
\batitle{{Stiffness design of geometrically nonlinear structures using topology
  optimization}}.
\bjtitle{Structural and Multidisciplinary Optimization}
\bvolume{19}(\bissue{2}),
\bfpage{93}--\blpage{104}
(\byear{2000}).
\doiurl{10.1007/s001580050089}
\end{barticle}
\endbibitem

%%% 20
\bibitem{Xue2022}
\begin{barticle}
\bauthor{\bsnm{Xue}, \binits{T.}},
\bauthor{\bsnm{Mao}, \binits{S.}}:
\batitle{{Mapped shape optimization method for the rational design of cellular
  mechanical metamaterials under large deformation}}.
\bjtitle{International Journal for Numerical Methods in Engineering}
\bvolume{123}(\bissue{10}),
\bfpage{2357}--\blpage{2380}
(\byear{2022}).
\doiurl{10.1002/nme.6941}
\end{barticle}
\endbibitem

%%% 21
\bibitem{Huang2022}
\begin{barticle}
\bauthor{\bsnm{Huang}, \binits{J.}},
\bauthor{\bsnm{Xu}, \binits{S.}},
\bauthor{\bsnm{Ma}, \binits{Y.}},
\bauthor{\bsnm{Liu}, \binits{J.}}:
\batitle{{A topology optimization method for hyperelastic porous structures
  subject to large deformation}}.
\bjtitle{International Journal of Mechanics and Materials in Design}
\bvolume{18}(\bissue{2}),
\bfpage{289}--\blpage{308}
(\byear{2022}).
\doiurl{10.1007/s10999-021-09576-4}
\end{barticle}
\endbibitem

%%% 22
\bibitem{Abdi2018}
\begin{barticle}
\bauthor{\bsnm{Abdi}, \binits{M.}},
\bauthor{\bsnm{Ashcroft}, \binits{I.}},
\bauthor{\bsnm{Wildman}, \binits{R.}}:
\batitle{{Topology optimization of geometrically nonlinear structures using an
  evolutionary optimization method}}.
\bjtitle{Engineering Optimization}
\bvolume{50}(\bissue{11}),
\bfpage{1850}--\blpage{1870}
(\byear{2018}).
\doiurl{10.1080/0305215X.2017.1418864}
\end{barticle}
\endbibitem

%%% 23
\bibitem{Kim2020}
\begin{barticle}
\bauthor{\bsnm{Kim}, \binits{S.}},
\bauthor{\bsnm{Yun}, \binits{G.J.}}:
\batitle{{Microstructure topology optimization by targeting prescribed
  nonlinear stress-strain relationships}}.
\bjtitle{International Journal of Plasticity}
\bvolume{128}(\bissue{January}),
\bfpage{102684}
(\byear{2020}).
\doiurl{10.1016/j.ijplas.2020.102684}
\end{barticle}
\endbibitem

%%% 24
\bibitem{Wang2022}
\begin{barticle}
\bauthor{\bsnm{Wang}, \binits{Y.}},
\bauthor{\bsnm{Zeng}, \binits{Q.}},
\bauthor{\bsnm{Wang}, \binits{J.}},
\bauthor{\bsnm{Li}, \binits{Y.}},
\bauthor{\bsnm{Fang}, \binits{D.}}:
\batitle{{Inverse design of shell-based mechanical metamaterial with customized
  loading curves based on machine learning and genetic algorithm}}.
\bjtitle{Computer Methods in Applied Mechanics and Engineering}
\bvolume{401},
\bfpage{115571}
(\byear{2022}).
\doiurl{10.1016/j.cma.2022.115571}
\end{barticle}
\endbibitem

%%% 25
\bibitem{Deng2022}
\begin{barticle}
\bauthor{\bsnm{Deng}, \binits{B.}},
\bauthor{\bsnm{Zareei}, \binits{A.}},
\bauthor{\bsnm{Ding}, \binits{X.}},
\bauthor{\bsnm{Weaver}, \binits{J.C.}},
\bauthor{\bsnm{Rycroft}, \binits{C.H.}},
\bauthor{\bsnm{Bertoldi}, \binits{K.}}:
\batitle{{Inverse Design of Mechanical Metamaterials with Target Nonlinear
  Response via a Neural Accelerated Evolution Strategy}}.
\bjtitle{Advanced Materials}
\bvolume{34}(\bissue{41}),
\bfpage{2206238}
(\byear{2022}).
\doiurl{10.1002/adma.202206238}
\end{barticle}
\endbibitem

%%% 26
\bibitem{Sohl-Dickstein2015}
\begin{botherref}
\oauthor{\bsnm{Sohl-Dickstein}, \binits{J.}},
\oauthor{\bsnm{Weiss}, \binits{E.A.}},
\oauthor{\bsnm{Maheswaranathan}, \binits{N.}},
\oauthor{\bsnm{Ganguli}, \binits{S.}}:
{Deep Unsupervised Learning using Nonequilibrium Thermodynamics}.
ICML
(2015)
{\href{https://arxiv.org/abs/1503.03585}{{arXiv:1503.03585}}}
\end{botherref}
\endbibitem

%%% 27
\bibitem{Ramesh2022}
\begin{botherref}
\oauthor{\bsnm{Ramesh}, \binits{A.}},
\oauthor{\bsnm{Dhariwal}, \binits{P.}},
\oauthor{\bsnm{Nichol}, \binits{A.}},
\oauthor{\bsnm{Chu}, \binits{C.}},
\oauthor{\bsnm{Chen}, \binits{M.}}:
{Hierarchical Text-Conditional Image Generation with CLIP Latents}
(2022)
{\href{https://arxiv.org/abs/2204.06125}{{arXiv:2204.06125}}}
\end{botherref}
\endbibitem

%%% 28
\bibitem{Ho2022}
\begin{botherref}
\oauthor{\bsnm{Ho}, \binits{J.}},
\oauthor{\bsnm{Chan}, \binits{W.}},
\oauthor{\bsnm{Saharia}, \binits{C.}},
\oauthor{\bsnm{Whang}, \binits{J.}},
\oauthor{\bsnm{Gao}, \binits{R.}},
\oauthor{\bsnm{Gritsenko}, \binits{A.}},
\oauthor{\bsnm{Kingma}, \binits{D.P.}},
\oauthor{\bsnm{Poole}, \binits{B.}},
\oauthor{\bsnm{Norouzi}, \binits{M.}},
\oauthor{\bsnm{Fleet}, \binits{D.J.}},
\oauthor{\bsnm{Salimans}, \binits{T.}}:
{Imagen Video: High Definition Video Generation with Diffusion Models},
90--91
(2022)
{\href{https://arxiv.org/abs/2210.02303}{{arXiv:2210.02303}}}
\end{botherref}
\endbibitem

%%% 29
\bibitem{Kingma2014}
\begin{botherref}
\oauthor{\bsnm{Kingma}, \binits{D.P.}},
\oauthor{\bsnm{Welling}, \binits{M.}}:
{Auto-Encoding Variational Bayes}.
2nd International Conference on Learning Representations, ICLR 2014 -
  Conference Track Proceedings
(Ml),
1--14
(2013)
{\href{https://arxiv.org/abs/1312.6114}{{arXiv:1312.6114}}}
\end{botherref}
\endbibitem

%%% 30
\bibitem{Goodfellow2014}
\begin{botherref}
\oauthor{\bsnm{Goodfellow}, \binits{I.J.}},
\oauthor{\bsnm{Pouget-Abadie}, \binits{J.}},
\oauthor{\bsnm{Mirza}, \binits{M.}},
\oauthor{\bsnm{Xu}, \binits{B.}},
\oauthor{\bsnm{Warde-Farley}, \binits{D.}},
\oauthor{\bsnm{Ozair}, \binits{S.}},
\oauthor{\bsnm{Courville}, \binits{A.}},
\oauthor{\bsnm{Bengio}, \binits{Y.}}:
{Generative Adversarial Networks}
(2014)
{\href{https://arxiv.org/abs/1406.2661}{{arXiv:1406.2661}}}
\end{botherref}
\endbibitem

%%% 31
\bibitem{Dhariwal2021}
\begin{botherref}
\oauthor{\bsnm{Dhariwal}, \binits{P.}},
\oauthor{\bsnm{Nichol}, \binits{A.}}:
{Diffusion Models Beat GANs on Image Synthesis}
(2021)
{\href{https://arxiv.org/abs/2105.05233}{{arXiv:2105.05233}}}
\end{botherref}
\endbibitem

%%% 32
\bibitem{Abueidda2019}
\begin{barticle}
\bauthor{\bsnm{Abueidda}, \binits{D.W.}},
\bauthor{\bsnm{Almasri}, \binits{M.}},
\bauthor{\bsnm{Ammourah}, \binits{R.}},
\bauthor{\bsnm{Ravaioli}, \binits{U.}},
\bauthor{\bsnm{Jasiuk}, \binits{I.M.}},
\bauthor{\bsnm{Sobh}, \binits{N.A.}}:
\batitle{{Prediction and optimization of mechanical properties of composites
  using convolutional neural networks}}.
\bjtitle{Composite Structures}
\bvolume{227},
\bfpage{111264}
(\byear{2019}).
\doiurl{10.1016/j.compstruct.2019.111264}
\end{barticle}
\endbibitem

%%% 33
\bibitem{Xiao2022}
\begin{barticle}
\bauthor{\bsnm{Xiao}, \binits{Y.}},
\bauthor{\bsnm{Hu}, \binits{D.}},
\bauthor{\bsnm{Zhang}, \binits{Z.}},
\bauthor{\bsnm{Pei}, \binits{B.}},
\bauthor{\bsnm{Wu}, \binits{X.}},
\bauthor{\bsnm{Lin}, \binits{P.}}:
\batitle{{A 3D-Printed Sole Design Bioinspired by Cat Paw Pad and Triply
  Periodic Minimal Surface for Improving Paratrooper Landing Protection}}.
\bjtitle{Polymers}
\bvolume{14}(\bissue{16}),
\bfpage{3270}
(\byear{2022}).
\doiurl{10.3390/polym14163270}
\end{barticle}
\endbibitem

%%% 34
\bibitem{Liu2020}
\begin{botherref}
\oauthor{\bsnm{Liu}, \binits{S.}},
\oauthor{\bsnm{Wang}, \binits{F.}},
\oauthor{\bsnm{Liu}, \binits{Z.}},
\oauthor{\bsnm{Zhang}, \binits{W.}},
\oauthor{\bsnm{Tian}, \binits{Y.}},
\oauthor{\bsnm{Zhang}, \binits{D.}}:
{A Two-Finger Soft-Robotic Gripper with Enveloping and Pinching Grasping
  Modes}.
IEEE/ASME Transactions on Mechatronics,
1--1
(2020).
\doiurl{10.1109/TMECH.2020.3005782}
\end{botherref}
\endbibitem

%%% 35
\bibitem{Jin2020}
\begin{barticle}
\bauthor{\bsnm{Jin}, \binits{L.}},
\bauthor{\bsnm{Khajehtourian}, \binits{R.}},
\bauthor{\bsnm{Mueller}, \binits{J.}},
\bauthor{\bsnm{Rafsanjani}, \binits{A.}},
\bauthor{\bsnm{Tournat}, \binits{V.}},
\bauthor{\bsnm{Bertoldi}, \binits{K.}},
\bauthor{\bsnm{Kochmann}, \binits{D.M.}}:
\batitle{{Guided transition waves in multistable mechanical metamaterials}}.
\bjtitle{Proceedings of the National Academy of Sciences}
\bvolume{117}(\bissue{5}),
\bfpage{2319}--\blpage{2325}
(\byear{2020}).
\doiurl{10.1073/pnas.1913228117}
\end{barticle}
\endbibitem

%%% 36
\bibitem{Nie2020}
\begin{botherref}
\oauthor{\bsnm{Nie}, \binits{Z.}},
\oauthor{\bsnm{Lin}, \binits{T.}},
\oauthor{\bsnm{Jiang}, \binits{H.}},
\oauthor{\bsnm{Kara}, \binits{L.B.}}:
{TopologyGAN: Topology Optimization Using Generative Adversarial Networks Based
  on Physical Fields Over the Initial Domain}.
Journal of Mechanical Design
\textbf{143}(3)
(2021)
{\href{https://arxiv.org/abs/2003.04685}{{arXiv:2003.04685}}}.
\doiurl{10.1115/1.4049533}
\end{botherref}
\endbibitem

%%% 37
\bibitem{Ho2020}
\begin{botherref}
\oauthor{\bsnm{Ho}, \binits{J.}},
\oauthor{\bsnm{Jain}, \binits{A.}},
\oauthor{\bsnm{Abbeel}, \binits{P.}}:
{Denoising Diffusion Probabilistic Models}.
Advances in Neural Information Processing Systems
(NeurIPS 2020),
1--25
(2020)
{\href{https://arxiv.org/abs/2006.11239}{{arXiv:2006.11239}}}
\end{botherref}
\endbibitem

%%% 38
\bibitem{Ho2021}
\begin{botherref}
\oauthor{\bsnm{Ho}, \binits{J.}},
\oauthor{\bsnm{Salimans}, \binits{T.}}:
{Classifier-Free Diffusion Guidance}
(2022)
{\href{https://arxiv.org/abs/2207.12598}{{arXiv:2207.12598}}}
\end{botherref}
\endbibitem

%%% 39
\bibitem{Ronneberger2015}
\begin{botherref}
\oauthor{\bsnm{Ronneberger}, \binits{O.}},
\oauthor{\bsnm{Fischer}, \binits{P.}},
\oauthor{\bsnm{Brox}, \binits{T.}}:
{U-Net: Convolutional Networks for Biomedical Image Segmentation}
(2015)
{\href{https://arxiv.org/abs/1505.04597}{{arXiv:1505.04597}}}
\end{botherref}
\endbibitem

%%% 40
\bibitem{Vlassis2023}
\begin{botherref}
\oauthor{\bsnm{Vlassis}, \binits{N.N.}},
\oauthor{\bsnm{Sun}, \binits{W.}}:
{Denoising diffusion algorithm for inverse design of microstructures with
  fine-tuned nonlinear material properties}
(2022),
1--21
(2023)
{\href{https://arxiv.org/abs/2302.12881}{{arXiv:2302.12881}}}
\end{botherref}
\endbibitem

%%% 41
\bibitem{Vaswani2017}
\begin{botherref}
\oauthor{\bsnm{Vaswani}, \binits{A.}},
\oauthor{\bsnm{Shazeer}, \binits{N.}},
\oauthor{\bsnm{Parmar}, \binits{N.}},
\oauthor{\bsnm{Uszkoreit}, \binits{J.}},
\oauthor{\bsnm{Jones}, \binits{L.}},
\oauthor{\bsnm{Gomez}, \binits{A.N.}},
\oauthor{\bsnm{Kaiser}, \binits{L.}},
\oauthor{\bsnm{Polosukhin}, \binits{I.}}:
{Attention Is All You Need}
(2017)
{\href{https://arxiv.org/abs/1706.03762}{{arXiv:1706.03762}}}
\end{botherref}
\endbibitem

%%% 42
\bibitem{Shaw2018}
\begin{botherref}
\oauthor{\bsnm{Shaw}, \binits{P.}},
\oauthor{\bsnm{Uszkoreit}, \binits{J.}},
\oauthor{\bsnm{Vaswani}, \binits{A.}}:
{Self-Attention with Relative Position Representations}
(2018)
{\href{https://arxiv.org/abs/1803.02155}{{arXiv:1803.02155}}}
\end{botherref}
\endbibitem

%%% 43
\bibitem{Firouzeh2017}
\begin{barticle}
\bauthor{\bsnm{Firouzeh}, \binits{A.}},
\bauthor{\bsnm{Salerno}, \binits{M.}},
\bauthor{\bsnm{Paik}, \binits{J.}}:
\batitle{{Stiffness Control With Shape Memory Polymer in Underactuated Robotic
  Origamis}}.
\bjtitle{IEEE Transactions on Robotics}
\bvolume{33}(\bissue{4}),
\bfpage{765}--\blpage{777}
(\byear{2017}).
\doiurl{10.1109/TRO.2017.2692266}
\end{barticle}
\endbibitem

%%% 44
\bibitem{Hu2023}
\begin{barticle}
\bauthor{\bsnm{Hu}, \binits{Y.}},
\bauthor{\bsnm{Kochmann}, \binits{D.M.}}:
\batitle{{Atomistic insight into three-dimensional twin embryo growth in Mg
  alloys}}.
\bjtitle{Journal of Materials Science}
\bvolume{58}(\bissue{9}),
\bfpage{3972}--\blpage{3995}
(\byear{2023}).
\doiurl{10.1007/s10853-023-08263-3}
\end{barticle}
\endbibitem

%%% 45
\bibitem{Lang2011}
\begin{botherref}
\oauthor{\bsnm{Lang}, \binits{A.}},
\oauthor{\bsnm{Potthoff}, \binits{J.}}:
{Fast simulation of Gaussian random fields}.
Monte Carlo Methods and Applications
\textbf{17}(3)
(2011).
\doiurl{10.1515/mcma.2011.009}
\end{botherref}
\endbibitem

%%% 46
\bibitem{Paszke2019}
\begin{botherref}
\oauthor{\bsnm{Paszke}, \binits{A.}},
\oauthor{\bsnm{Gross}, \binits{S.}},
\oauthor{\bsnm{Massa}, \binits{F.}},
\oauthor{\bsnm{Lerer}, \binits{A.}},
\oauthor{\bsnm{Bradbury}, \binits{J.}},
\oauthor{\bsnm{Chanan}, \binits{G.}},
\oauthor{\bsnm{Killeen}, \binits{T.}},
\oauthor{\bsnm{Lin}, \binits{Z.}},
\oauthor{\bsnm{Gimelshein}, \binits{N.}},
\oauthor{\bsnm{Antiga}, \binits{L.}},
\oauthor{\bsnm{Desmaison}, \binits{A.}},
\oauthor{\bsnm{K{\"{o}}pf}, \binits{A.}},
\oauthor{\bsnm{Yang}, \binits{E.}},
\oauthor{\bsnm{DeVito}, \binits{Z.}},
\oauthor{\bsnm{Raison}, \binits{M.}},
\oauthor{\bsnm{Tejani}, \binits{A.}},
\oauthor{\bsnm{Chilamkurthy}, \binits{S.}},
\oauthor{\bsnm{Steiner}, \binits{B.}},
\oauthor{\bsnm{Fang}, \binits{L.}},
\oauthor{\bsnm{Bai}, \binits{J.}},
\oauthor{\bsnm{Chintala}, \binits{S.}}:
{PyTorch: An imperative style, high-performance deep learning library}.
Advances in Neural Information Processing Systems
(2019)
{\href{https://arxiv.org/abs/1912.01703}{{arXiv:1912.01703}}}
\end{botherref}
\endbibitem

%%% 47
\bibitem{Wang2022a}
\begin{botherref}
\oauthor{\bsnm{Wang}, \binits{P.}}:
{Implementation of Imagen, Google's Text-to-Image Neural Network that beats
  DALL-E2, in Pytorch}
(2022).
\url{https://github.com/lucidrains/imagen-pytorch}
\end{botherref}
\endbibitem

%%% 48
\bibitem{Zagoruyko2016}
\begin{botherref}
\oauthor{\bsnm{Zagoruyko}, \binits{S.}},
\oauthor{\bsnm{Komodakis}, \binits{N.}}:
{Wide Residual Networks}
(2016)
{\href{https://arxiv.org/abs/1605.07146}{{arXiv:1605.07146}}}
\end{botherref}
\endbibitem

%%% 49
\bibitem{Elfwing2017}
\begin{botherref}
\oauthor{\bsnm{Elfwing}, \binits{S.}},
\oauthor{\bsnm{Uchibe}, \binits{E.}},
\oauthor{\bsnm{Doya}, \binits{K.}}:
{Sigmoid-Weighted Linear Units for Neural Network Function Approximation in
  Reinforcement Learning}
(2017)
{\href{https://arxiv.org/abs/1702.03118}{{arXiv:1702.03118}}}
\end{botherref}
\endbibitem

%%% 50
\bibitem{Katharopoulos2020}
\begin{botherref}
\oauthor{\bsnm{Katharopoulos}, \binits{A.}},
\oauthor{\bsnm{Vyas}, \binits{A.}},
\oauthor{\bsnm{Pappas}, \binits{N.}},
\oauthor{\bsnm{Fleuret}, \binits{F.}}:
{Transformers are RNNs: Fast Autoregressive Transformers with Linear Attention}
(2020)
{\href{https://arxiv.org/abs/2006.16236}{{arXiv:2006.16236}}}
\end{botherref}
\endbibitem

\end{thebibliography}


\end{document}